\documentclass[conference]{IEEEtran}
\IEEEoverridecommandlockouts
\usepackage{cite}
\usepackage{amsmath,amssymb,amsfonts}
\usepackage{algorithmic}
\usepackage{graphicx}
\usepackage{textcomp}
\usepackage{xcolor}
\usepackage{hyperref}
\usepackage{enumitem}
\usepackage[caption=false,font=footnotesize]{subfig}

\def\BibTeX{{\rm B\kern-.05em{\sc i\kern-.025em b}\kern-.08em
    T\kern-.1667em\lower.7ex\hbox{E}\kern-.125emX}}

\begin{document}

\title{Practical Insights on Incremental Learning of New Human Physical Activity on the Edge
}
\author{\IEEEauthorblockN{George Arvanitakis, Jingwei Zuo, Mthandazo Ndhlovu and Hakim Hacid}
\IEEEauthorblockA{Technology Innovation Institute, Abu Dhabi, UAE \\
Email: $\{$firstname.lastname$\}$@tii.ae}
}

\maketitle

\begin{abstract}
Edge Machine Learning (Edge ML), which shifts computational intelligence from cloud-based systems to edge devices, is attracting significant interest due to its evident benefits including reduced latency, enhanced data privacy, and decreased connectivity reliance. While these advantages are compelling, they introduce unique challenges absent in traditional cloud-based approaches.
In this paper, we delve into the intricacies of Edge-based learning, examining the interdependencies among: (i) constrained data storage on Edge devices, (ii) limited computational power for training, and (iii) the number of learning classes. Through experiments conducted using our MAGNETO system, that focused on learning human activities via data collected from mobile sensors, we highlight these challenges and offer valuable perspectives on Edge ML.
\end{abstract}

\begin{IEEEkeywords}
 Edge ML, Human Activity Recognition, Dynamic Class Integration, Incremental Learning
\end{IEEEkeywords}

%

\section{Introduction}

The proliferation of Internet of Things (IoT) has led to an exploding demand for network resources. Furthermore, ensuring the security and privacy of users' data is becoming paramount across a multitude of applications. The paradigm of Edge Machine Learning (Edge ML), seeks to address these challenges by pushing  ML pipelines to Edge devices~\cite{Murshed_2022}.
One of the intuitive use cases where IoT sensors intersect with ML is Human Activity Recognition (HAR). This involves leveraging sensors from everyday commercial devices to deduce a user's activities.
In fact, this case epitomizes many of the characteristics and constraints of the field, underscoring the imperative for ML on the Edge. 
Traditional approaches for predicting human physical activity predominantly rely on training a classifier on a predefined set of activity classes in a centralized cloud environment. Subsequently, user measurements captured on devices are then sent to the cloud for inference~\cite{ANTAR2021146}. However, this centralized, cloud-based learning approach suffers from three main drawbacks: high latency due to user-cloud communication, lack of flexibility and personalization to individual user's needs, and lower privacy control as user's raw data is consistently relayed over networks to the cloud.

In contrast with the conventional ML scenarios, where the main processing is performed on remote cloud servers even when applications run on local devices, Edge ML~\cite{Lee2018} brings the core processing tasks to the edge devices. This approach facilitates the deployment of optimized models and services directly onto user devices or the edge network, ensuring rapid real-time response, low latency, offline capability, and enhanced security and privacy. 
%

%
However, moving the inference, or more ambitiously, the learning process to the {Edge} devices introduces a host of significant challenges, stem primarily from the inherent limitations of Edge devices, including (i) Model size, which should be small enough to fit within the {Edge} but also to operate efficiently,  (ii) Data size, which should be very limited due to the low storage capabilities within the {Edge}, and (iii) Energy consumption, constraining the training process to be very efficient without excessive power consumption. 
Regarding on-device inference, recent advancements in {Edge ML} have been notable. Many studies~\cite{agarwal2020lightweight, 10.1145/3457388.3458656} emphasize human activity recognition on smart devices by training light ML models in terms of memory and complexity. While there's interest in flexible ML models~\cite{zuo2019isets} that can learn new human activities, like those using few-shots learning~\cite{FENG2019112782, s20030825}, these often overlook that unique constraints of the edge constraints.

This paper introduces the essential elements of our contrastive learning methodology for building and continually updating ML models directly on the Edge. It demonstrates the feasibility of incrementally learning new classes on the fly directly on the Edge. Through our result analysis, we provide practical insights into the performance and the technical constraints that may govern this sensitive task of learning on the Edge, e.g., learning with limited observations or sequentially learning more than one task. 
The rest of this paper is organized as follows: 
Section~\ref{sec:magnetosystem} presents the MAGNETO system for human activity recognition at a high level. Section~\ref{sec:problemformulation} discusses in more detail the different dimensions of the problem and the modeling assumptions. Section~\ref{sec:experiments} reviews the results and provides practical insights about learning on the Edge. Finally, we conclude and provide some future work in Section~\ref{sec:conclusion}.


\section{The Magneto System}
\label{sec:magnetosystem}
Smart devices, including smartphones and smartwatches, use built-in sensors to detect and predict user activities. This area has garnered substantial interest recently, leading to numerous studies and datasets like the Huawei-Sussex locomotion challenge~\cite{8418369} and the Transportation Mode Detection dataset~\cite{8480119}. Big tech firms, such as Google\footnote{\url{https://developers.google.com/location-context/activity-recognition}}, Samsung, and Apple\footnote{\url{https://developer.apple.com/documentation/coremotion/cmmotionactivity}}, have integrated these capabilities. Notably, most research and applications have predominantly adopted centralized or cloud-based approaches.


MAGNETO, \textit{sMArt sensinG for humaN activity rEcogniTiOn}, is an implemented system that provides human activity recognition via sensor measurements from ordinary, commercial, smart devices (e.g., smartphones and smartwatches). Despite the large body of literature, MAGNETO provides inference of human activity on an Edge device, using a pre-trained model, without transferring user’s data to the cloud. 

Furthermore, MAGNETO is equipped with the ability to incrementally learn new activities, i.e., classes, by capturing extra user data in order to (i) re-calibrate an activity to be more accurate for user's personal style or (ii) re-train the model to learn a custom new activity according to user’s habits, without any data exchange with the cloud. We believe that the activity recognition on the Edge combined with the capability of learning new actions relying on personalized needs can enable a new area in the health care, fitness or assistant applications. In the next section, we present in detail the approach and the followed nuances of the inference and learning on the Edge.

\section{System Architecture}
\label{sec:problemformulation}
\begin{figure}[h]
    \centering
    \includegraphics[width=0.92\linewidth]{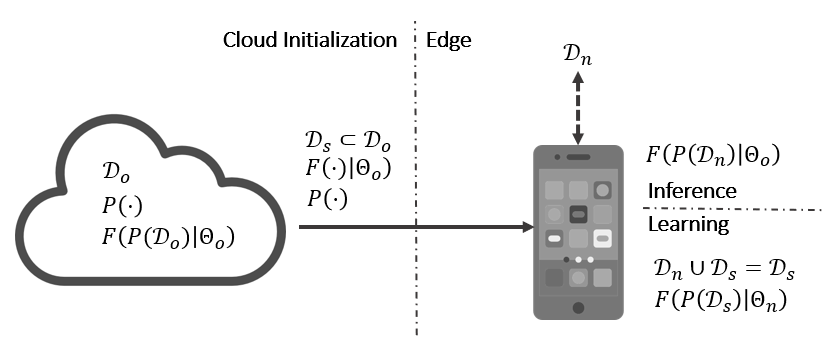}
    \caption{Illustration of the proposed architecture, showing the dependencies between Cloud and Edge for model's inference and learning}
    \label{fig:architecture}
\end{figure}

This section presents all the mandatory components of the system architecture (Figure \ref{fig:architecture}) as well as their interconnections. The overall process can be split into two phases (i) cloud initialization, which aims at pre-training our model in order to avoid the usually high data and power demand of initial models construction, and after transfers to the Edge device all the functions and data that are mandatory for inference and learning on the Edge. ii) Inference and Learning on the Edge, which operates all the mandatory actions for inference and learning new activities on the Edge without any exchange of user's data with the cloud.

\subsection{Cloud initialisation }
\label{sec:edgeinference}

The main components of the cloud initialization are: 

\textbf{Initial data set}, $\mathcal{D}_o$: Those data, representing $K$ initial classes, are stored in the cloud for our model's initial training, using activity data from our measurement campaign.

\textbf{The pre-processing function}, $P(\cdot)$: This function processes raw sensor data to prepare it for the {ML} algorithm. In our implementation, the pre-processing function takes roughly 120 sequential measurements from 22 mobile sensors, such as accelerometers and gyroscopes, over a one-second period. It then calculates statistics like average, variance, average/variance of the jerk for each feature sensor. In total, a set of 86 features are extracted to represent the activity.

\textbf{The Initial  {ML} Model}, $F(\cdot|\Theta_o)$: using all the available data $\mathcal{D}_o$ and the preprocessing function $P(\cdot)$, a small neural network $F\cdot|\Theta_o$ of dimensions [$1024 \times 512 \times 128 \times 64 \times 64 $], with a contrastive loss cost function~\cite{khosla2020supervised} and ADAM optimizer is trained on the cloud. The choice of the contrastive loss is motivated by its ease of adaptation and ability to learn on new tasks with a very limited amount of data~\cite{Koch2015SiameseNN}.

\textbf{The support set} $\mathcal{D}_s$: For Edge learning, a foundational set of observations is essential to facilitate the learning process, whether it's to create data pairs for contrastive or triplet loss, or to leverage "old" data to prevent catastrophic forgetting~\cite{zuo2023handling}). This is termed the support set $\mathcal{D}_s$. Comprising a fraction of data samples from each class, $\mathcal{D}_s$ is significantly smaller than the original dataset and is a subset of it, represented as: $|\mathcal{D}_s| << |\mathcal{D}_o| $ and $\mathcal{D}_s \subset \mathcal{D}_o$. The size of the support set is pivotal to the learning process. Additionally, the exact selection of the support set can influence the quality of the learning process. However, determining the optimal makeup of the support set, while crucial, is beyond this paper's scope.

\textbf{Prototypes} $P_o$: Using the support set and the trained model, we calculate the prototypes of each class, on the latent space, that are used for inference.

At the end of the initialization phase, the Web cloud passes on the Edge device three mandatory items: (i) pre-processing function, (ii) the Initial ML model, and (iii) the support set.

\begin{figure*}
\centering
\setkeys{Gin}{width=0.32\linewidth}
\subfloat[Initial model trained without the \textit{Run} activity]{\includegraphics[width=0.31\linewidth]{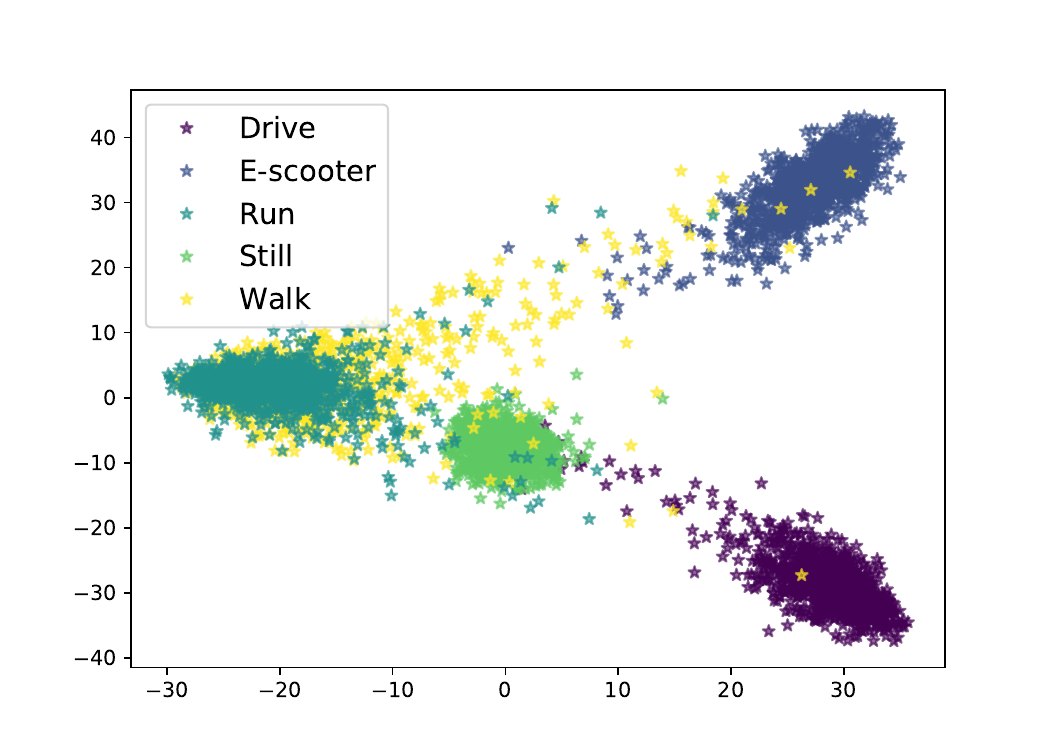}\label{fig:performance:1}}
\hfill
\subfloat[Re-training 10 epochs, using 200 \textit{Run} samples]{\includegraphics[width=0.31\linewidth]{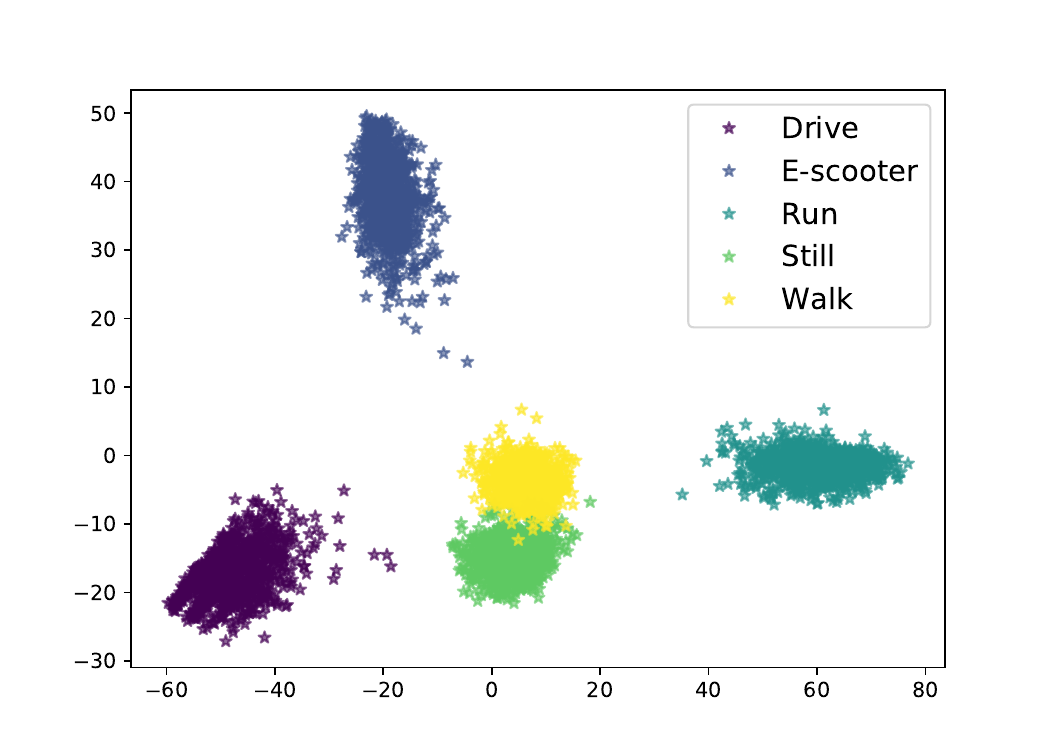}\label{fig:performance:2}}
\hfill
\subfloat[Overall performance w.r.t. the support set size]{\includegraphics{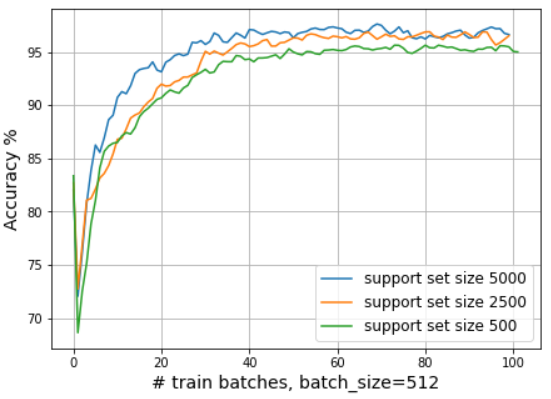}\label{fig:performance:3}}
\hfill
\subfloat[Learning performance of each of new classes]{\includegraphics{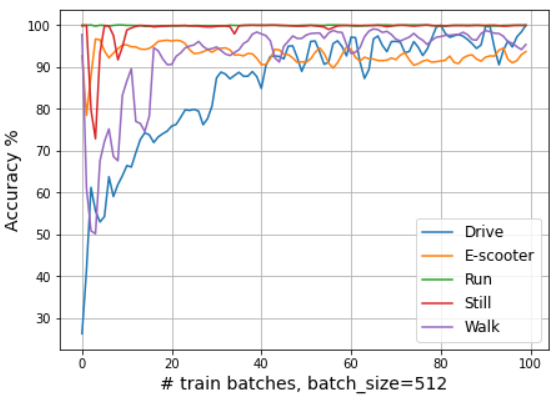}\label{fig:performance:4}}
\hfill
\subfloat[Average learning performance of the old classes while learning the new one]{\includegraphics{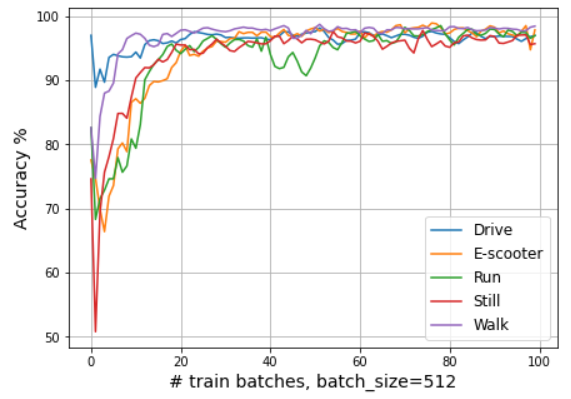}\label{fig:performance:5}}
\hfill
\subfloat[Overall performance of all the classes]{\includegraphics{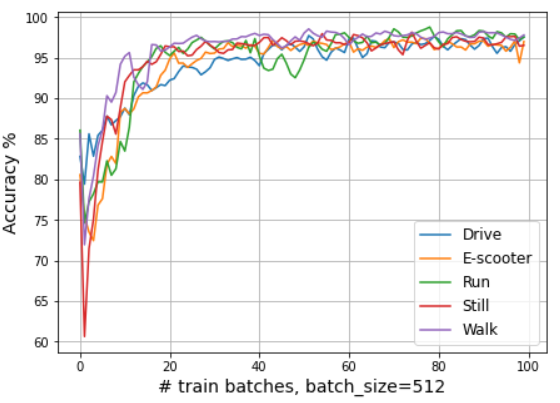}\label{fig:performance:6}}

\caption{Performance results for single class learning on the Edge}

\end{figure*}

\subsection{Inference and Learning on the  {Edge}}

\subsubsection{Inference} Using the transferred components, the Edge device is able to infer user's activity on the fly by reading its sensors and passing the captured measurements sequentially from the pre-processing function $P(\cdot)$ to $F(\cdot | \Theta_o)$ and compare the output representation with the classes prototypes.
\subsubsection{Learning new activities on the Edge} 
the processing steps that take place on the {Edge} device are as follows: 
\begin{itemize}[leftmargin=1.5em]

    \item \textbf{Samples Collection} $\mathcal{D}_n$: The user records samples of a new activity not present in the initial dataset. This new annotated data is integrated into the Edge device's existing support set $\mathcal{D}_s = \mathcal{D}_s \cup \mathcal{D}_n$.

    \item \textbf{Model Re-training} $F(\cdot | \Theta_n)$: The extended support set is used to retrain the existing model, expanding the learned class count from $K$ to $K+1$.

    \item \textbf{Prototype Update} $P_n$: With the new support set and the re-trained model, the class prototypes are updated.
\end{itemize}

It's noteworthy that the process for \textit{re-calibration} of an existing class (tailored to a user's behavior) mirrors the one described earlier. The primary difference being, in re-calibration, the current activity data in the support set is swapped with new data, followed by retraining the model on the same activities.


\section{Experiments and Insights}
\label{sec:experiments}
In this section, we detail the experiments performed on a real-world dataset. These experiments not only demonstrate the viability of our proposed method but also shed light on the primary constraints influencing Edge-based learning.


\subsection{Experimental Set-up}

In experiments used our human activity dataset $\mathcal{D}$, consisted of more than 100GB of sensory data, annotated for 5 human activities \textit{(Drive, E-scooter, Run, Still, Walk)}, that comes from our data collection campaign. The dataset is split into training and test sets  $\mathcal{D} =  \{\mathcal{D}_{tr},  \mathcal{D}_{ts}\}$. To make both subsets totally disjoint, we ensure that there is a time distance of at least 10 seconds between the training and the test observations.

We selected accuracy and convergence speed as our performance metrics for the learning process. Convergence speed is determined by counting the number of training samples (batch-wise) required for the algorithm to approach its final accuracy within a margin of $±2\%$.
As mentioned before, we employed a five-layer fully connected network with a contrastive loss and Adam optimizer. Each training batch includes 512 sample pairs from the designated training set, i.e., batch size is set to 512.
The experimental process is the following:    
\begin{enumerate}[leftmargin=1.5em]
    \item From $\mathcal{D}_{tr}$, one activity is excluded, forming the \textit{initial} dataset $\mathcal{D}_o$ with $K=4$ activities. This dataset is used to train $F(\cdot | \Theta_o)$.
    \item A subset of $\mathcal{D}_o$ is randomly sampled for each class as a support set, $\mathcal{D}_s$.
    \item A part of the excluded class is used as observations of the new class $\mathcal{D}_n$. To eliminate any bias, the amount of the new observations equals to the amount of the $K$ existing classes that are included in the support set, i.e., $|\mathcal{D}_n| =|\mathcal{D}_s|/K$.
    \item Using the support set and new observations, the model $F(\cdot | \Theta_n)$ is re-trained.
    \item \textit{Step 2} is repeated with varying support set sizes: $|\mathcal{D}_s| = [1000, 500, 100]$ per class.
    \item \textit{Steps 1-5} are reiterated five times for each class.

\end{enumerate}

\subsection{Performance Analysis}
In this section, we discuss the different results with a focus on three specific areas: (i) the \textit{feasibility of incremental learning} on the Edge, (ii) the \textit{impact of the support size} on the quality of the learning, and (ii) the performance of \textit{learning multiple classes} sequentially. 

\subsubsection{Learning Performance with respect to the new activity} At first, we infer the new class data in the concept of zero-shot learning, without re-training the initial model $F(\cdot | \Theta_o)$.
Without any extra training of the model, the accuracy of the new class is almost random choice $~20\%$. Figure~\ref{fig:performance:1} shows the embedding space of the initial model without training on the \textit{Run} activity. The model is mainly confusing \textit{Run} with the \textit{Walk} activity. However, Figure~\ref{fig:performance:2} shows the updated embedding space after a brief re-training of just 10 epochs using 200 \textit{Run} samples. This brief adaptation greatly enhances the distinction of the new class. Hence, while zero-shot learning doesn't align with theoretical expectations, even a limited re-training with minimal data from the new class can effectuate significant improvement on model's performance.

To delve deeper into how learning a new class affects the performance of both the new and existing classes, we show in Figure~\ref{fig:performance:4}, \ref{fig:performance:5}, \ref{fig:performance:6} the model's performance on different metrics per each new class, considering $|D_{s}|=1000$. Notably, certain new classes, such as \textit{Still}, are learned faster than others (see Figure~\ref{fig:performance:4}). Although, a broader perspective reveals that the \textit{overall learning performance} of all the classes is converging similarly (see Figure~\ref{fig:performance:6}). Moreover, it's remarkable to note that the learning does not demand extensive iterations to converge ($\approx$40 batches). This agile convergence persists across varying the size of the support set (see Figure~\ref{fig:performance:3}). 

The outcome shows clearly the possibility of operating incremental learning on the Edge despite the limited resources. 

\subsubsection{Learning Performance Vs Support set size}

The next important question arises concerning storage capacity required by the Edge device to hold the support set. The support set is a critical component for incremental learning as historical data is needed to reconstruct or refine models, it is important to discern the bounds and potentials of this parameter. 
Figure \ref{fig:performance:3} shows the average performance across all classes, with varying sizes of the support set: $[5000, 2500, 500]$ total observations.

Two salient observations emerge from this: i) The size of the support set seems to not have an important effect on the \textit{convergence speed}. All cases converge around the $40^{th}$ training batch; ii) There is a slight degradation in \textit{accuracy} (less than 2\%) between the case of $|\mathcal{D}_s| =5000$ and $|\mathcal{D}_s| =500$. This marginal trade-off, however, offers tangible benefits: the support set's size decreased by a factor of ten, leading to diminished processing demands and a lighter storage footprint. Both these facets are particularly advantageous within the Edge environment, given its inherent resource constraints. 

\subsubsection{Learning Performance Vs Multiple new classes}

\begin{figure}
    \centering
    \includegraphics[width=0.9\linewidth]{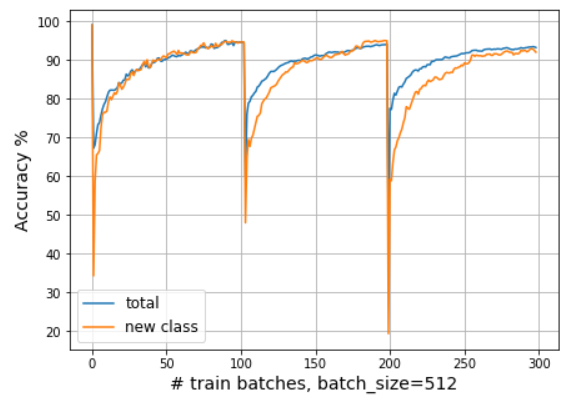}
    \caption{Overall performance for adding sequential tasks.  Accuracy in every stage: Initial (2 activities) = 98.9, +1 activity = 96, +2 activities = 94.5 and +3 activities = 93.1}
    \label{fig:multiacc}
\end{figure}

The last question has to do with the feasibility of sequentially learning multiple activities directly on the Edge.
To provide an answer to this question, an initial dataset that includes only two activities is first used to build the starting model. After that, the model is sequentially trained over the other three, i.e., one by one. To ensure a robust analysis, we generate 100 random realizations of activity combinations. For instance,  one particular sequence might involve kick-starting with an initial set of \textit{(Drive, Walk)}, thereafter learning the \textit{Still} activity first, followed by \textit{E-scooter} and concluding with \textit{Run}. We report the model's average performance. 
Figure~\ref{fig:multiacc} shows the aggregated accuracy from all the realizations using a support set of 1000 observations per class. 

The obtained results expose interesting outcomes: First, the sequential learning of multiple classes (activities) on the Edge is doable. This is particularly important for the Edge context where users dynamically increment their tasks. Second, there is a degradation of roughly $30\%$ at the beginning. This could be explained by the fact that the initial model was trained on the simple task of learning 2 activities and has less generalization capabilities. Third, as new activities are learned, the degradation exists, but the overall rate is slow, maintaining very good overall performances of the system. Finally, the convergence of the newly added classes seem to become slower as we move forward in the sequence. 


\section{Conclusion}
\label{sec:conclusion}
This paper explored and demonstrated the feasibility of incremental learning on the Edge for human activity recognition. We highlighted interesting findings around the impact of different factors, including the support set size and training epochs. Furthermore, our research showcased the effects of sequentially introducing multiple new classes. As for future work, one can consider the scenario with more classes, and explore more intricate machine learning tasks on the Edge.



\bibliographystyle{ieeetr}
\bibliography{biblio}

\end{document}